\documentclass[12pt,thmsa]{article}
\usepackage{sw20lart}

\input{tcilatex}
\begin{document}

\title{Exclusive $B\rightarrow K_{1}\ell ^{+}\ell ^{-}$ decay in model with
single universal extra dimension}
\author{Ishtiaq Ahmed, M. Ali Paracha \\
Department of Physics and National Centre for Physics, \\
Quaid-i-Azam University, Islamabad, Pakistan. \and M. Jamil\ Aslam \\
Department of Physics,\\
COMSATS Institute of Information Technology, Islamabad, Pakistan.}
\maketitle

\begin{abstract}
Decay rate and forward-backward asymmetries in $B\rightarrow K_{1}\ell
^{+}\ell ^{-}$ , $K_{1}$ is the axial vector meson, are calculated in the
universal extra dimension (UED) model. The dependence of these physical
quantities on the compactification radius $R$, the only unknown paramter in
UED\ model, is studied and it is shown that zero of forward-backward
asymmetry is sensitive to the UED\ model, therefore they can be very useful
tool to establish new physics predicted by the UED\ model. This work is
briefly extended to $B\rightarrow K^{*}l^{+}l^{-}.$
\end{abstract}

\section{Introduction}

The flavor-changing-neutral-current (FCNC) transitions $b\rightarrow s$
provide potentially stringent tests of standard model (SM) in flavor physics
and are not allowed at tree level but are induced by the
Glashow-Iliopoulos-Miani (GIM) amplitudes \cite{1} at the loop level in the
SM. In addition, these are also suppressed in SM due to their dependence on
the weak mixing angles of the quark-flavor rotation matrix $-$ the
Cabibbo-Kobayashi-Maskawa (CKM) matrix \cite{2}. These two circumstances
make the FCNC decays relatively rare and hence important for the study of
physics beyond the SM commonly known as new physics.

The experimental observation of inclusive \cite{3} and exclusive \cite{4}
decays, $B\rightarrow X_{s}\gamma $ and $B\rightarrow K^{\ast }\gamma $ ,
has prompted a lot of theoretical interest on rare $B$ meson decays. Though
the inclusive decays are theoretically better understood but are extremely
difficult to be measured in a hadron mechine, such as the LHC, which is the
only collider, except for a Super-$B$ factory, that could provide enough
luminosity for the precise study of the decay distribution of such rare
processes. In contrast, the exclusive decays are easy to detect
experimentally but are challanging to calculate theoretically and the
difficulty lies in describing the hadronic structure, which provides the
main uncertainty in the predictions of exclusive rare decays. In exclusive $%
B\rightarrow K,K^{\ast }$ decays the long-distance effects in the meson
transition amplitude of the effective Hamiltonian are encoded in the meson
transition form factors which are the scalar functions of the square of
momentum transfer and are model dependent quantites. Many exclusive $%
B\rightarrow K\left( K^{\ast }\right) \ell ^{+}\ell ^{-}$\cite{10,11,12}, $%
B\rightarrow \phi \ell ^{+}\ell ^{-}$\cite{13a}, $B\rightarrow \gamma \ell
^{+}\ell ^{-}$\cite{13}, $B\rightarrow \ell ^{+}\ell ^{-}$\cite{14}
processes based on $b\rightarrow s\left( d\right) \ell ^{+}\ell ^{-}$ have
been studied in literature and many fameworks have been applied to the
description of meson transition form factors: like constituent quark models,
QCD sum rules, lattice QCD, approaches based on heavy quark symmetry and
analytical constraints.

Rare $B$ decay modes also provide imporatant ways to look for physics beyond
the SM. There are various extensions of the SM in the literature, but the
models with extra dimensions are of viable interest as they provide a
unified framework for gravity and other interactions. In this way they give
some hints on the hierarchy problem and a connection with string theory.
Among different models of extra dimensions, which differ from one another
depending on the number of extra dimensions, the most interesting are the
scenario with universal extra dimensions. In these UED models all the SM
fields are allowed to propagate in the extra dimensions and compactification
of extra dimension leads to the appearance of Kaluza-Klein (KK) partners of
the SM fields in the four dimensional description of higher dimensional
theory, together with KK modes without corresponding SM partners. The
Appelquist, Cheng and Dobrescu (ACD) model \cite{15} with one universal
extra dimension (UED) is very attractive because it has only one free
parameter with respect to the SM and that is the inverse of compactification
radius $R$ \cite{16}.

By analyzing the signature of extra dimenions in the different processes,
one can get bounds to the size of extra dimensions which are different in
different models. These bounds are accessible for the processes already
known at the particle accelerators or within the reach of planned future
facilities. In case of UED these bounds are more sever and constraints from
Tevatron run I allow to put the bound $1/R\geq 300$ GeV \cite{16}.

Rare $B$ decays can also be used to constraint the ACD scenario and in this
regard Buras and collaborators have already done some work. In addition to
the effective Hamiltonian they have calculated for $b-s$ decays and also
investigated the impact of UED on the $B^{0}-\bar{B}^{0}$ mixing as well as
on the CKM unitarity triangle \cite{17, 18, 19}. Due to the availability of
precise data on the decays $B\rightarrow K\left( K^{*}\right) \ell ^{+}\ell
^{-}$, Colangelo et al. have studied these decays in ACD model by
calculating the branching ratio and forward-backward asymmetry for the decay 
$B\rightarrow K\left( K^{*}\right) \ell ^{+}\ell ^{-}$. We will study the
rare semileptonic decay modes, $B\rightarrow K_{1}\ell ^{+}\ell ^{-}$on the
same footing as $B\rightarrow K^{*}\ell ^{+}\ell ^{-}$ because both are
induced by the same quark level transitions, i.e. $b\rightarrow s\ell
^{+}\ell ^{-}$. We compare results of forward backward asymmetry for $%
B\rightarrow K^{*}l^{+}l^{-}$ using our form factors with those obtained by
Colangelo $etal$ \cite{16}. The comparision shows clear distinction as shown
in Fig 4. These decays may provide us step forward towards the study of
existance of new physics beyond the SM and therefore deserve serious
attention, both theoretically and experimentally.

The paper is organized as follows. In Section 2 we will briefly introduce
the ACD model. Section 3 deals with the study of effective Hamiltonian and
the corresponding matrix elements for $B\rightarrow K_{1}\ell ^{+}\ell ^{-}$
decay. Now the new physics manifest in these decays in two different ways,
either through new operators in the in the effective Hamiltonian which are
absent in the SM or through new contributions to the Wilson coefficients 
\cite{13a}. In ACD no new operator appears at tree level and therefore the
new physics comes only through the Wilson coefficients which are calculated
in literature \cite{18, 19} and we will summarize them in the same section.
Finally, in Section 4 we will calculate the decay rate and forward-backward
asymmetry and summarize our results.

\section{ACD Model}

In our usual universe we have 3 spatial $+$1 temporal dimensions and if an
extra dimension exists and is compactified, fields living in all dimensions
would menifest themselves in the $3+1$ space by the appearence of
Kaluza-Klein excitations. The most pertinent question is whether ordinary
fields propagate or not in all extra dimensions. One obvious possibilty is
the propagation of gravity in whole ordinary plus extra dimensional
universe, the ``bulk''. Contrary to this there are the models with universal
extra dimensions (UED) in which all the fields propagate in all available
dimensions \cite{15} and Appelquist, Cheng and Dobrescu model belongs to one
of UED scenarios \cite{16}

This model is the minimal extension of the SM in $4+\delta $ dimensions, and
in literature a simple case $\delta =1$ is considered \cite{16}. The
topology for this extra dimension is orbifold $S^{1}/Z_{2}$, and the
coordinate $x_{5}=y$ runs from $0$ to $2\pi R$, where $R$ is the the
compactification radius. The Kaluza-Klein (KK) mode expension of the fields
are determined from the boundary conditions at two fixed points $y=0$ and $%
y=\pi R$ on the orbifold. Under parity transformation $P_{5}:$ $y\rightarrow
-y$ the fields may be even or odd. Even fields have their correspondent in
the $4$ dimensional SM and their zero mode in the KK mode expansion can be
interpreted as the ordionary SM field. The odd fields do not have their
correspondent in the SM and therefore do not have zero mode in the KK
expansion.

The significant features of the ACD model are:

\begin{itemize}
\item[i)] the compactification radius $R$ is the only free parameter with
respect to SM

\item[ii)] no tree level contribution of KK modes in low energy processes
(at scale $\mu \ll 1/R$) and no production of single KK excitation in
ordinary particle interactions is a consequence of conservation of KK parity.
\end{itemize}

The detailed description of ACD model is provided in \cite{18}; here we
summarize main features of its construction from \cite{16}.

\textbf{Gauge group}

As ACD model is the minimal extension of SM therefore the gauge bosons
associated with the gauge group $SU\left( 2\right) _{L}\times U\left(
1\right) _{Y}$ are $W_{i}^{a}\,(a=1,\,2,\,3$, $i=0$,$\,1$,$\,2$,$\,3$,$\,5)$
and $B_{i}$, and the gauge couplings are $\hat{g}_{2}=g_{2}\sqrt{2\pi R}$
and $\hat{g}^{\prime }=g^{\prime }\sqrt{2\pi R}$ (the hat on the coupling
constant refers to the extra dimension). The charged bosons are $W_{i}^{\pm
}=\frac{1}{\sqrt{2}}\left( W_{i}^{1}\mp W_{i}^{2}\right) $ and the mixing of 
$W_{i}^{3}$ and $B_{i}$ give rise to the fields $Z_{i}$ and $A_{i}$ as they
do in the SM. The relations for the mixing angles are: 
\begin{equation}
c_{W}=\cos \theta _{W}=\frac{\hat{g}_{2}}{\sqrt{\hat{g}_{2}^{2}+\hat{g}%
^{\prime 2}}}\,\,\,\,\,\,\,\,\,\,\,\,\,\,\,\,\,\,\,\,\,c_{W}=\sin \theta
_{W}=\frac{\hat{g}^{\prime }}{\sqrt{\hat{g}_{2}^{2}+\hat{g}^{\prime 2}}}\,
\label{couplings}
\end{equation}
The Weingberg angle remains the same as in the SM, due to the relationship
between five and four dimensional constants. The gluons which are the gauge
bosons associated to $SU\left( 3\right) _{C}$ are $G_{i}^{a}\left(
x,y\right) (a=1,\ldots ,8)$.

\textbf{Higgs sector and mixing between Higgs fields and gauge bosons}

The Higgs doublet can be written as: 
\begin{equation}
\phi =\left( 
\begin{array}{l}
i\chi ^{+} \\ 
\frac{1}{\sqrt{2}}\left( \psi -i\chi ^{3}\right)%
\end{array}
\right)  \label{higgs-doublet}
\end{equation}
with $\chi ^{\pm }=\frac{1}{\sqrt{2}}\left( \chi ^{1}\mp \chi ^{2}\right) $.
Now only field $\psi $ has a zero mode, and we assign vacuum expectation
value $\hat{v}$ to such mode, so that $\psi \rightarrow \hat{v}+H$. $H$ is
the the SM Higgs field, and the relation between expectation values in five
and four dimension is: $\hat{v}=v/\sqrt{2\pi R}$.

The Goldstone fields $G_{\left( n\right) }^{0}$, $G_{\left( n\right) }^{\pm
} $ aries due to the mixing of charged $W_{5\left( n\right) }^{\pm }$ and $%
\chi _{\left( n\right) }^{\pm }$ , as well as neutral fields $Z_{5\left(
n\right) }$. These Goldstone modes are then used to give masses to the $%
W_{\left( n\right) }^{\pm \mu }$ and $Z_{\left( n\right) }^{\mu }$, and $%
a_{\left( n\right) }^{0}$, $a_{\left( n\right) }^{\pm }$, new physical
scalars.

\textbf{Yukawa terms}

In SM, Yukawa coupling of the Higgs field to the fermion provides the
fermion mass terms. The diagonalization of such terms leads to the
introduction of the CKM matrix. In order to have chiral fermions in ACD
model, the left and right-handed components of the given spinor cannot be
simultaneously even under $P_{5}$.\ This makes the ACD model to be the
minimal flavor violation model, since there are no new opeators beyond those
present in the SM and no new phase beyond the CKM phase and the unitarity
triangle remains the same as in\ SM \cite{18}. In order to have 4-d mass
eigenstates of higher KK levels, a further mixing is introduced among the
left-handed doublet and right-handed singlet of each flavor $f$. The mixing
angle is such that $\tan \left( 2\alpha _{f\left( n\right) }\right) =\frac{%
m_{f}}{n/R}\left( n\geq 1\right) $ giving mass $m_{f\left( n\right) }=\sqrt{%
m_{f}^{2}+\frac{n^{2}}{R^{2}}}$, so that it is negligible for all flavors
except the top \cite{16}.

Integrating over the fifth-dimension $y$ gives the four-dimensional
Lagrangian: 
\begin{equation}
\mathcal{L}_{4}\left( x\right) =\int_{0}^{2\pi R}\mathcal{L}_{5}\left(
x,y\right)  \label{4-dllagrangian}
\end{equation}
which describes: (i) zero modes corresponding to the SM\ fields, (ii) their
massive KK excitations, (iii) KK excitations without zero modes which do not
corresponds to any field in SM. Feynman rules used in the further
calculation are given in Ref. \cite{18}.

\section{Effective Hamiltonian}

At quark level the decay $B\rightarrow K_{1}\ell ^{+}\ell ^{-}$ is same like 
$B\rightarrow K^{\ast }\ell ^{+}\ell ^{-}$ as discussed by Ali \textit{et al}%
.\cite{10}, i.e.$b\rightarrow s\ell ^{+}\ell ^{-}$ and it can be described
by effective Hamiltonian obtained by integrating out the top quark and $%
W^{\pm }$ bosons 
\begin{equation}
H_{eff}=-4\frac{G_{F}}{\sqrt{2}}V_{tb}V_{ts}^{\ast }\sum_{i=1}^{10}C_{i}(\mu
)O_{i}(\mu )  \label{2.1}
\end{equation}
where $O_{i}$'$s$ are four local quark operators and $\dot{C}_{i}$ are
Wilson co-effeicents calculated in Naive dimensional regularization (NDR)
scheme \cite{21}.

One can write the above Hamiltonian in the following free quark decay
amplitude 
\begin{eqnarray}
\mathcal{M}(b &\rightarrow &s\ell ^{+}\ell ^{-})=\frac{G_{F}\alpha }{\sqrt{2}%
\pi }V_{tb}V_{ts}^{\ast }\left\{ 
\begin{array}{c}
C_{9}^{eff}\left[ \bar{s}\gamma _{\mu }Lb\right] \left[ \bar{\ell}\gamma
^{\mu }\ell \right] \\ 
+C_{10}\left[ \bar{s}\gamma _{\mu }Lb\right] \left[ \bar{\ell}\gamma ^{\mu
}\gamma ^{5}\ell \right] \\ 
-2\hat{m}_{b}C_{7}^{eff}\left[ \bar{s}i\sigma _{\mu \nu }\frac{\hat{q}^{\nu }%
}{\hat{s}}Rb\right] \left[ \bar{\ell}\gamma ^{\mu }\ell \right]%
\end{array}
\right\}  \nonumber \\
&&  \label{2.2}
\end{eqnarray}
with $L/R\equiv \frac{\left( 1\mp \gamma _{5}\right) }{2}$, $s=q^{2}$ which
is just the momentum transfer from heavy to light meson. The amplitude given
in Eq. (\ref{2.2}) contains long distance effects encoded in the form
factors and short distance effects that are hidden in Wilson coefficients.
These Wilson coeffients have been computed at next-to-next leading order
(NNLO) in the SM \cite{22}. Specifically for exclusive decays, the effective
coefficient $C_{9}^{eff}$ can be written as 
\begin{equation}
C_{9}^{eff}=C_{9}+Y\left( \hat{s}\right)  \label{resonances1}
\end{equation}
where the perturbatively calculated result of $Y\left( \hat{s}\right) $ is 
\cite{21} 
\begin{equation}
Y_{\text{pert}}\left( \hat{s}\right) =\left. 
\begin{array}{c}
g\left( \hat{m}_{c}\text{,}\hat{s}\right) \left(
3C_{1}+C_{2}+3C_{3}+C_{4}+3C_{5}+C_{6}\right) \\ 
-\frac{1}{2}g\left( 1\text{,}\hat{s}\right) \left(
4C_{3}+4C_{4}+3C_{5}+C_{6}\right) \\ 
-\frac{1}{2}g\left( 0\text{,}\hat{s}\right) \left( C_{3}+3C_{4}\right) +%
\frac{2}{9}\left( 3C_{3}+C_{4}+3C_{5}+C_{6}\right) .%
\end{array}
\right.  \label{reson-expr}
\end{equation}
Here the hat denote the normalization in term of $B$ meson mass. For the
explicit expressions of $g$'s and numerical values of the Wilson
coefficients appearing in Eq. (\ref{reson-expr}) we refer to \cite{21}.

In ACD model the new physics comes through the Wilson coefficients. Buras et
al. have computed the above coefficients at NLO in ACD model including the
effects of KK modes \cite{18, 19}; we use these results to study $%
B\rightarrow K_{1}\ell ^{+}\ell ^{-}$ decay. As it has already been
mentioned that ACD model is the minimal extension of SM with only one extra
dimension and it has no extra operator other than the SM, therefore, the
whole contribution from all the KK states is in the Wilson coefficients,
i.e. now they depend on the additional ACD parameter, the inverse of
compactification radius $R$. At large value of $1/R$ the SM\ phenomenology
should be recovered, since the new states, being more and more massive,
decoupled from the low-energy theory. Our objective is to calculate the
decay rate and forward-backward asymmetry for $B\rightarrow K_{1}\ell
^{+}\ell ^{-}$ using the lower bound on $1/R$ provided by Colangelo et al.
for $B\rightarrow K^{\ast }\ell ^{+}\ell ^{-}$ decay \cite{16}.

In ACD model, the Wilson coefficients are modified and they contain the
contribution from new particles which are not present in the SM and comes as
an intermediate state in penguin and box diagrams. Thus, these coefficients
can be expressed in terms of the functions $F\left( x_{t},1/R\right) $, $%
x_{t}=\frac{m_{t}^{2}}{M_{W}^{2}}$, which generalize the corresponding SM\
function $F_{0}\left( x_{t}\right) $ according to: 
\begin{equation}
F\left( x_{t},1/R\right) =F_{0}\left( x_{t}\right) +\tsum_{n=1}^{\infty
}F_{n}\left( x_{t},x_{n}\right)  \label{f-expression}
\end{equation}
with $x_{n}=\frac{m_{n}^{2}}{M_{W}^{2}}$ and $m_{n}=\frac{n}{R}$ \cite{16}.
The relevant diagrams are $Z^{0}$ penguins, $\gamma $ penguins, gluon
penguins, $\gamma $ magnetic penguins, Chormomagnetic penguins$\ $and the
corresponding functions are $C\left( x_{t},1/R\right) $, $D\left(
x_{t},1/R\right) $, $E\left( x_{t},1/R\right) $, $D^{\prime }\left(
x_{t},1/R\right) $ and $E^{\prime }\left( x_{t},1/R\right) $ respectively.
These functions can be found in \cite{18, 19} but to make the paper self
contained, we collect here the formulae needed for our analysis.

$\bullet C_{7}$

In place of $C_{7},$ one defines an effective coefficient $C_{7}^{(0)eff}$
which is renormalization scheme independent \cite{23}: 
\begin{equation}
C_{7}^{(0)eff}(\mu _{b})=\eta ^{\frac{16}{23}}C_{7}^{(0)}(\mu _{w})+\frac{8}{%
3}(\eta ^{\frac{14}{23}}-\eta ^{\frac{16}{23}})C_{8}^{(0)}(\mu
_{w})+C_{2}^{(0)}(\mu _{w})\sum_{i=1}^{8}h_{i}\eta ^{\alpha _{i}}
\label{wilson1}
\end{equation}
where $\eta =\frac{\alpha s(\mu _{w})}{\alpha _{s}(\mu _{b})},$ and 
\begin{equation}
C_{2}^{(0)}(\mu _{w})=1,\text{ }C_{7}^{(0)}(\mu _{w})=-\frac{1}{2}D^{\prime
}(x_{t},\frac{1}{R}),\text{ }C_{8}^{(0)}(\mu _{w})=-\frac{1}{2}E^{\prime
}(x_{t},\frac{1}{R});  \label{wilson2}
\end{equation}
the superscript $(0)$ stays for leading $\log $ approximation. Furthermore: 
\begin{eqnarray}
\alpha _{1} &=&\frac{14}{23}\text{ \quad }\alpha _{2}=\frac{16}{23}\text{
\quad }\alpha _{3}=\frac{6}{23}\text{ \quad }\alpha _{4}=-\frac{12}{23} 
\nonumber \\
\alpha _{5} &=&0.4086\text{ \quad }\alpha _{6}=-0.4230\text{ \quad }\alpha
_{7}=-0.8994\text{ \quad }\alpha _{8}=-0.1456  \nonumber \\
h_{1} &=&2.996\text{ \quad }h_{2}=-1.0880\text{ \quad }h_{3}=-\frac{3}{7}%
\text{ \quad }h_{4}=-\frac{1}{14}  \nonumber \\
h_{5} &=&-0.649\text{ \quad }h_{6}=-0.0380\text{ \quad }h_{7}=-0.0185\text{
\quad }h_{8}=-0.0057.  \label{wilson3}
\end{eqnarray}
The functions $D^{\prime }$ and $E^{\prime }$ are given be eq. (\ref{wilson3}%
) with 
\begin{equation}
D_{0}^{\prime }(x_{t})=-\frac{(8x_{t}^{3}+5x_{t}^{2}-7x_{t})}{12(1-x_{t})^{3}%
}+\frac{x_{t}^{2}(2-3x_{t})}{2(1-x_{t})^{4}}\ln x_{t}  \label{wilson4}
\end{equation}
\begin{equation}
E_{0}^{\prime }(x_{t})=-\frac{x_{t}(x_{t}^{2}-5x_{t}-2)}{4(1-x_{t})^{3}}+%
\frac{3x_{t}^{2}}{2(1-x_{t})^{4}}\ln x_{t}  \label{wilson5}
\end{equation}
\begin{eqnarray}
D_{n}^{\prime }(x_{t},x_{n}) &=&\frac{%
x_{t}(-37+44x_{t}+17x_{t}^{2}+6x_{n}^{2}(10-9x_{t}+3x_{t}^{2})-3x_{n}(21-54x_{t}+17x_{t}^{2}))%
}{36(x_{t}-1)^{3}}  \nonumber \\
&&+\frac{x_{n}(2-7x_{n}+3x_{n}^{2})}{6}\ln \frac{x_{n}}{1+x_{n}}  \nonumber
\\
&&-\frac{%
(-2+x_{n}+3x_{t})(x_{t}+3x_{t}^{2}+x_{n}^{2}(3+x_{t})-x_{n})(1+(-10+x_{t})x_{t}))%
}{6(x_{t}-1)^{4}}\ln \frac{x_{n}+x_{t}}{1+x_{n}}  \label{wilson6}
\end{eqnarray}
\begin{eqnarray}
E_{n}^{\prime }(x_{t},x_{n}) &=&\frac{%
x_{t}(-17-8x_{t}+x_{t}^{2}+3x_{n}(21-6x_{t}+x_{t}^{2})-6x_{n}^{2}(10-9x_{t}+3x_{t}^{2}))%
}{12(x_{t}-1)^{3}}  \nonumber \\
&&+-\frac{1}{2}x_{n}(1+x_{n})(-1+3x_{n})\ln \frac{x_{n}}{1+x_{n}}  \nonumber
\\
&&+\frac{%
(1+x_{n})(x_{t}+3x_{t}^{2}+x_{n}^{2}(3+x_{t})-x_{n}(1+(-10+x_{t})x_{t}))}{%
2(x_{t}-1)^{4}}\ln \frac{x_{n}+x_{t}}{1+x_{n}}  \label{wilson7}
\end{eqnarray}
Following \cite{18} one gets the expressions for the sum over $n:$%
\begin{eqnarray}
\sum_{n=1}^{\infty }D_{n}^{\prime }(x_{t},x_{n}) &=&-\frac{%
x_{t}(-37+x_{t}(44+17x_{t}))}{72(x_{t}-1)^{3}}  \nonumber \\
&&+\frac{\pi M_{w}R}{2}[\int_{0}^{1}dy\frac{2y^{\frac{1}{2}}+7y^{\frac{3}{2}%
}+3y^{\frac{5}{2}}}{6}]\coth (\pi M_{w}R\sqrt{y})  \nonumber \\
&&+\frac{(-2+x_{t})x_{t}(1+3x_{t})}{6(x_{t}-1)^{4}}J(R,-\frac{1}{2}) 
\nonumber \\
&&-\frac{1}{6(x_{t}-1)^{4}}%
[x_{t}(1+3x_{t})-(-2+3x_{t})(1+(-10+x_{t})x_{t})]J(R,\frac{1}{2})  \nonumber
\\
&&+\frac{1}{6(x_{t}-1)^{4}}[(-2+3x_{t})(3+x_{t})-(1+(-10+x_{t})x_{t})]J(R,%
\frac{3}{2})  \nonumber \\
&&-\frac{(3+x_{t})}{6(x_{t}-1)^{4}}J(R,\frac{5}{2})]  \label{wilson8}
\end{eqnarray}
\begin{eqnarray}
\sum_{n=1}^{\infty }E_{n}^{\prime }(x_{t},x_{n}) &=&-\frac{%
x_{t}(-17+(-8+x_{t})x_{t})}{24(x_{t}-1)^{3}}  \nonumber \\
&&+\frac{\pi M_{w}R}{2}[\int_{0}^{1}dy(y^{\frac{1}{2}}+2y^{\frac{3}{2}}-3y^{%
\frac{5}{2}})\coth (\pi M_{w}R\sqrt{y})]  \nonumber \\
&&-\frac{x_{t}(1+3x_{t})}{(x_{t}-1)^{4}}J(R,-\frac{1}{2})  \nonumber \\
&&+\frac{1}{(x_{t}-1)^{4}}[x_{t}(1+3x_{t})-(1+(-10+x_{t})x_{t})]J(R,\frac{1}{%
2})  \nonumber \\
&&-\frac{1}{(x_{t}-1)^{4}}[(3+x_{t})-(1+(-10+x_{t})x_{t})]J(R,\frac{3}{2}) 
\nonumber \\
&&+\frac{(3+x_{t})}{(x_{t}-1)^{4}}J(R,\frac{5}{2})]  \label{wilson9}
\end{eqnarray}
where 
\begin{equation}
J(R,\alpha )=\int_{0}^{1}dyy^{\alpha }[\coth (\pi M_{w}R\sqrt{y}%
)-x_{t}^{1+\alpha }\coth (\pi m_{t}R\sqrt{y})].  \label{wilson10}
\end{equation}
$\bullet C_{9}$

In the ACD model and in the NDR scheme one has 
\begin{equation}
C_{9}(\mu )=P_{0}^{NDR}+\frac{Y(x_{t},\frac{1}{R})}{\sin ^{2}\theta _{W}}%
-4Z(x_{t},\frac{1}{R})+P_{E}E(x_{t},\frac{1}{R})  \label{wilson11}
\end{equation}
where $P_{0}^{NDR}=2.60\pm 0.25[20]$ and the last term is numerically
negligible. Besides 
\begin{eqnarray}
Y(x_{t},\frac{1}{R}) &=&Y_{0}(x_{t})+\sum_{n=1}^{\infty }C_{n}(x_{t},x_{n}) 
\nonumber \\
Z(x_{t},\frac{1}{R}) &=&Z_{0}(x_{t})+\sum_{n=1}^{\infty }C_{n}(x_{t},x_{n})
\label{wilson12}
\end{eqnarray}
with 
\begin{eqnarray}
Y_{0}(x_{t}) &=&\frac{x_{t}}{8}[\frac{x_{t}-4}{x_{t}-1}+\frac{3x_{t}}{%
(x_{t}-1)^{2}}\ln x_{t}]  \nonumber \\
Z_{0}(x_{t}) &=&\frac{18x_{t}^{4}-163x_{t}^{3}+259x_{t}^{2}-108x_{t}}{%
144(x_{t}-1)^{3}}  \nonumber \\
&&+[\frac{32x_{t}^{4}-38x_{t}^{3}+15x_{t}^{2}-18x_{t}}{72(x_{t}-1)^{4}}-%
\frac{1}{9}]\ln x_{t}  \label{wilson13}
\end{eqnarray}
\begin{equation}
C_{n}(x_{t},x_{n})=\frac{x_{t}}{8(x_{t}-1)^{2}}%
[x_{t}^{2}-8x_{t}+7+(3+3x_{t}+7x_{n}-x_{t}x_{n})\ln \frac{x_{t}+x_{n}}{%
1+x_{n}}]  \label{wilson14}
\end{equation}
and 
\begin{equation}
\sum_{n=1}^{\infty }C_{n}(x_{t},x_{n})=\frac{x_{t}(7-x_{t})}{16(x_{t}-1)}-%
\frac{\pi M_{w}Rx_{t}}{16(x_{t}-1)^{2}}[3(1+x_{t})J(R,-\frac{1}{2}%
)+(x_{t}-7)J(R,\frac{1}{2})]  \label{wilson15}
\end{equation}
$\bullet C_{10}$

$C_{10}$ is $\mu $ independent and is given by 
\begin{equation}
C_{10}=-\frac{Y(x_{t},\frac{1}{R})}{\sin ^{2}\theta _{w}}.  \label{wilson16}
\end{equation}
The normalization scale is fixed to $\mu =\mu _{b}\simeq 5$ GeV.

Wilson coefficients give the short distance effects where as the long
distance effects involve the matrix elements of the operators in Eq. (\ref%
{2.2}) between the $B$ and $K_{1}$ mesons. Using standard parameterization
in terms of the form factors we have \cite{24}: 
\begin{eqnarray}
\left\langle K_{1}(k,\varepsilon )\left| V_{\mu }\right| B(p)\right\rangle
&=&i\varepsilon _{\mu }^{\ast }\left( M_{B}+M_{K_{1}}\right) V_{1}(s) 
\nonumber \\
&&-(p+k)_{\mu }\left( \varepsilon ^{\ast }\cdot q\right) \frac{V_{2}(s)}{%
M_{B}+M_{K_{1}}}  \nonumber \\
&&-q_{\mu }\left( \varepsilon \cdot q\right) \frac{2M_{K_{1}}}{s}\left[
V_{3}(s)-V_{0}(s)\right]  \label{matrix-1} \\
\left\langle K_{1}(k,\varepsilon )\left| A_{\mu }\right| B(p)\right\rangle
&=&\frac{2i\epsilon _{\mu \nu \alpha \beta }}{M_{B}+M_{K_{1}}}\varepsilon
^{\ast \nu }p^{\alpha }k^{\beta }A(s)  \label{matrix2}
\end{eqnarray}
where $V_{\mu }=\bar{s}\gamma _{\mu }b$ and $A_{\mu }=\bar{s}\gamma _{\mu
}\gamma _{5}b$ are the vector and axial vector currents respectively and $%
\varepsilon _{\mu }^{\ast }$ is the polarization vector for the final state
axial vector meson.

The relationship between different form factors which also ensures that
there is no kinematical singularity in the matrix element at $s=0$ is

\begin{eqnarray}
V_{3}(s) &=&\frac{M_{B}+M_{K_{1}}}{2M_{K_{1}}}V_{1}(s)-\frac{M_{B}-M_{K_{1}}%
}{2M_{K_{1}}}V_{2}(s)  \label{matrix3} \\
V_{3}(0) &=&V_{0}(0).  \label{matrix4}
\end{eqnarray}
In addition to the above form factors there are also some penguin form
factors which are: 
\begin{eqnarray}
\left\langle K_{1}(k,\varepsilon )\left| \bar{s}i\sigma _{\mu \nu }q^{\nu
}b\right| B(p)\right\rangle &=&\left[ \left( M_{B}^{2}-M_{K_{1}}^{2}\right)
\varepsilon _{\mu }-(\varepsilon \cdot q)(p+k)_{\mu }\right] F_{2}(s) 
\nonumber \\
&&+(\varepsilon ^{\ast }\cdot q)\left[ q_{\mu }-\frac{s}{%
M_{B}^{2}-M_{K_{1}}^{2}}(p+k)_{\mu }\right] F_{3}(s)  \nonumber \\
&&  \label{matrix-5} \\
\left\langle K_{1}(k,\varepsilon )\left| \bar{s}i\sigma _{\mu \nu }q^{\nu
}\gamma _{5}b\right| B(p)\right\rangle &=&-i\epsilon _{\mu \nu \alpha \beta
}\varepsilon ^{\ast \nu }p^{\alpha }k^{\beta }F_{1}(s)  \label{matrix-6}
\end{eqnarray}
with $F_{1}(0)=2F_{2}(0).$

Form factors are the non-perturbative quantities and are the scalar function
of the square of momentum transfer. Different models are used to calculate
these form factors. The form factors we use here in the analysis of physical
variables like decay rate and forward-backward asymmetry have been
calculated using Ward identities. The detailed calculation and their
expressions are given in ref. \cite{24} and can be summarized as: 
\begin{eqnarray}
A\left( s\right) &=&\frac{A\left( 0\right) }{\left( 1-s/M_{B}^{2}\right)
(1-s/M_{B}^{\prime 2})}  \nonumber \\
V_{1}(s) &=&\frac{V_{1}(0)}{\left( 1-s/M_{B_{A}^{*}}^{2}\right) \left(
1-s/M_{B_{A}^{*}}^{\prime 2}\right) }\left( 1-\frac{s}{%
M_{B}^{2}-M_{K_{1}}^{2}}\right)  \label{form-factors} \\
V_{2}(s) &=&\frac{\tilde{V}_{2}(0)}{\left( 1-s/M_{B_{A}^{*}}^{2}\right)
\left( 1-s/M_{B_{A}^{*}}^{\prime 2}\right) }-\frac{2M_{K_{1}}}{%
M_{B}-M_{K_{1}}}\frac{V_{0}(0)}{\left( 1-s/M_{B}^{2}\right) \left(
1-s/M_{B}^{\prime 2}\right) }  \nonumber
\end{eqnarray}
with 
\begin{eqnarray}
A(0) &=&-(0.52\pm 0.05)  \nonumber \\
V_{1}(0) &=&-(0.24\pm 0.02)  \nonumber \\
\tilde{V}_{2}(0) &=&-(0.39\pm 0.03)  \label{Num-f-factor}
\end{eqnarray}
The corresponding values for $B\rightarrow K^{*}$ form factors at $s=0$ are
given by 
\begin{eqnarray}
V(0) &=&(0.29\pm 0.04)  \nonumber \\
A_{1}(0) &=&(0.23\pm 0.03)  \label{Kstformfactor} \\
\tilde{A}_{2}(0) &=&(0.33\pm 0.05)  \nonumber
\end{eqnarray}

\section{Decay Distribution and Forward-Backward Asymmetry}

In this section we define the decay rate distribution which we shall use for
the phenomenological analysis . Following the notation from ref.\cite{10} we
can write from Eq. (\ref{2.2}) 
\begin{equation}
\mathcal{M}=\frac{G_{F}\alpha }{2\sqrt{2}\pi }V_{tb}V_{ts}^{\ast }m_{B}\left[
\mathcal{T}_{\mu }^{1}\left( \bar{l}\gamma ^{\mu }l\right) +\mathcal{T}_{\mu
}^{2}\left( \bar{l}\gamma ^{\mu }\gamma ^{5}l\right) \right]  \label{4.1}
\end{equation}
where 
\begin{eqnarray}
\mathcal{T}_{\mu }^{1} &=&A\left( \hat{s}\right) \varepsilon _{\mu \rho
\alpha \beta }\epsilon ^{\ast \rho }\hat{p}_{B}^{\alpha }\hat{p}%
_{K_{1}}^{\beta }-iB\left( \hat{s}\right) \epsilon _{\mu }^{\ast }+iC\left( 
\hat{s}\right) \left( \epsilon ^{\ast }\cdot \hat{p}_{B}\right) \hat{p}%
_{h\mu }+iD\left( \hat{s}\right) \left( \epsilon ^{\ast }\cdot \hat{p}%
_{B}\right) \hat{q}_{\mu }  \nonumber \\
&&  \label{4.2} \\
\mathcal{T}_{\mu }^{2} &=&E\left( \hat{s}\right) \varepsilon _{\mu \rho
\alpha \beta }\epsilon ^{\ast \rho }\hat{p}_{B}^{\alpha }\hat{p}%
_{K_{1}}^{\beta }-iF\left( \hat{s}\right) \epsilon _{\mu }^{\ast }+iG\left( 
\hat{s}\right) \left( \epsilon ^{\ast }\cdot \hat{p}_{B}\right) \hat{p}%
_{h\mu }+iH\left( \hat{s}\right) \left( \epsilon ^{\ast }\cdot \hat{p}%
_{B}\right) \hat{q}_{\mu }  \nonumber \\
&&  \label{4.3}
\end{eqnarray}
The definition of different momenta involved are defined in reference\cite%
{10}, where the auxiliary functions are 
\begin{eqnarray}
A(\hat{s}) &=&-\frac{2A(\hat{s})}{1+\hat{M}_{K_{1}}}C_{9}^{eff}(\hat{s})+%
\frac{2\hat{m}_{b}}{\hat{s}}C_{7}^{eff}F_{1}(\hat{s})  \nonumber \\
B(\hat{s}) &=&\left( 1+\hat{M}_{K_{1}}\right) \left[ C_{9}^{eff}(\hat{s}%
)V_{1}(\hat{s})+\frac{2\hat{m}_{b}}{\hat{s}}C_{7}^{eff}\left( 1-\hat{M}%
_{K_{1}}\right) \right]  \nonumber \\
C\left( \hat{s}\right) &=&\frac{1}{\left( 1-\hat{M}_{K_{1}}^{2}\right) }%
\left\{ C_{9}^{eff}(\hat{s})V_{2}(\hat{s})+2\hat{m}_{b}C_{7}^{eff}\left[
F_{3}(\hat{s})+\frac{1-\hat{M}_{K_{1}}^{2}}{\hat{s}}F_{2}(\hat{s})\right]
\right\}  \nonumber \\
D(\hat{s}) &=&\frac{1}{\hat{s}}\left[ 
\begin{array}{c}
\left( C_{9}^{eff}(\hat{s})(1+\hat{M}_{K_{1}})V_{1}(\hat{s})-(1-\hat{M}%
_{K_{1}})V_{2}(\hat{s})-2\hat{M}_{K_{1}}V_{0}(\hat{s})\right) \\ 
-2\hat{m}_{b}C_{7}^{eff}F_{3}(\hat{s})%
\end{array}
\right]  \nonumber \\
E(\hat{s}) &=&-\frac{2A(\hat{s})}{1+\hat{M}_{K_{1}}}C_{10}  \nonumber \\
F(\hat{s}) &=&\left( 1+\hat{M}_{K_{1}}\right) C_{10}V_{1}(\hat{s})  \nonumber
\\
G(\hat{s}) &=&\frac{1}{1+\hat{M}_{K_{1}}}C_{10}V_{2}(\hat{s})  \nonumber \\
H(\hat{s}) &=&\frac{1}{\hat{s}}\left[ C_{10}(\hat{s})(1+\hat{M}%
_{K_{1}})V_{1}(\hat{s})-(1-\hat{M}_{K_{1}})V_{2}(\hat{s})-2\hat{M}%
_{K_{1}}V_{0}(\hat{s})\right] .  \label{4.4}
\end{eqnarray}
Considering the final state lepton as muon the branching ratio for $%
B\rightarrow K_{1}\mu ^{+}\mu ^{-}$ is calculated in ref. \cite{24} and its
numerical value is 
\[
\mathcal{B}\left( B\rightarrow K_{1}\mu ^{+}\mu ^{-}\right)
=0.9_{-\,0.14}^{+\,0.11}\times 10^{-7} 
\]
\newline
The above value of branching ratio is for the case if one does not include $%
Y(\hat{s})$ in Eq. (\ref{reson-expr}). The error in the value reflects the
uncertainty from the form factors, and due to the variation of input
parameters like CKM matrix elements, decay constant of $B$ meson and masses
as defined in Table I.

\begin{eqnarray*}
\text{Table I} &:&\text{Default value of input parameters used in the
calculation} \\
&& \\
&& 
\begin{tabular}{ll}
\hline
$m_{W}$ & $80.41$ GeV \\ 
$m_{Z}$ & $91.1867$ GeV \\ 
$sin^{2}\theta _{W}$ & $0.2233$ \\ 
$m_{c}$ & $1.4$ GeV \\ 
$m_{b,pole}$ & $4.8\pm 0.2$ GeV \\ 
$m_{t}$ & $173.8\pm 5.0$ GeV \\ 
$\alpha _{s}\left( m_{Z}\right) $ & $0.119\pm 0.0058$ \\ 
$f_{B}$ & $\left( 200\pm 30\right) $ MeV \\ 
$\left| V_{ts}^{*}V_{tb}\right| $ & $0.0385$ \\ \hline
\end{tabular}%
\end{eqnarray*}

By including $Y(\hat{s})$ the central value of branching ratio reduces to 
\[
\mathcal{B}\left( B\rightarrow K_{1}\mu ^{+}\mu ^{-}\right) =0.72\times
10^{-7}
\]%
It is already mentioned that in ACD model there is no new opeartor beyond
the SM and new physics will come only through the Wilson coefficients. To
see this, the differential branching ratio against $\hat{s}$ is plotted in
Fig. 1 using the central values of input parameteres. One can see that their
is significant enhancement in the decay rate due to KK contribution for $%
1/R=200$ GeV whereas the value is shifted towards the SM at large value of $%
1/R$. The enhancemen is prominent in the low value of $\hat{s}$ but such
effects are obscured by the uncertainites involved in different parameters
like, form factors, CKM matrix elements etc. The numerical value at these
two different values of $1/R$ is 
\begin{eqnarray*}
\mathcal{B}\left( B\rightarrow K_{1}\mu ^{+}\mu ^{-}\right)  &=&0.82\times
10^{-7}\text{ for }1/R=200\text{ GeV} \\
\mathcal{B}\left( B\rightarrow K_{1}\mu ^{+}\mu ^{-}\right)  &=&0.75\times
10^{-7}\text{ for }1/R=500\text{ GeV}
\end{eqnarray*}%
The effects of UED becomes more clear if we look for the FB asymmetry in the
dilepton angular distribution because it depends upon the Wilson
coefficients. It is known that in SM, due to the opposite sign of the $C_{7}$
and $C_{9}$, the forward-backward asymmetry passes from its zero position
and has very weak dependence on form factors and uncertainities in input
parameteres. The differential forward-backward asymmetry for $B\rightarrow
K_{1}\mu ^{+}\mu ^{-}$ reads as follows\cite{10} 
\begin{equation}
\frac{d\mathcal{A}_{\text{FB}}}{d\hat{s}}=\frac{G_{F}^{2}\alpha ^{2}m_{B}^{5}%
}{2^{10}\pi ^{5}}\left\vert V_{ts}^{\ast }V_{tb}\right\vert ^{2}\hat{s}\hat{u%
}\left( \hat{s}\right) \left[ \text{Re}\left( BE^{\ast }\right) +\text{Re}%
\left( AF^{\ast }\right) \right]   \label{4.6}
\end{equation}%
where 
\begin{eqnarray}
\hat{u}\left( \hat{s}\right)  &=&\sqrt{\lambda \left( 1-4\frac{\hat{m}%
_{l}^{2}}{\hat{s}}\right) }  \nonumber \\
\lambda  &\equiv &\lambda \left( 1,\hat{m}_{K_{1}}^{2},\hat{s}\right)  
\nonumber \\
&=&1+\hat{m}_{K_{1}}^{4}+\hat{s}^{2}-2\hat{s}-2\hat{m}_{K_{1}}^{2}\left( 1+%
\hat{s}\right)   \label{4.7}
\end{eqnarray}%
The variable $\hat{u}$ corresponds to $\theta $, the angle between the
momentum of the $B$ meson and the positively charged lepton in the dilepton
c.m. system frame. In\ SM the zero-point of forward-backward asymmetry for $%
B\rightarrow K_{1}\mu ^{+}\mu ^{-}$ is \ calculated by Paracha et al. \cite%
{24} and it lies at $\hat{s}=0.16$ $\left( s=4.46\text{ GeV}^{-2}\right) $.
They have shown that the due to the uncertainities in the form factors zero
position of forward-backward asymmetry $A_{FB}$ deviate slightly from the
central value in the low $s$ region where as in the large $s$ region these
deviations are highly suppressed and zero of the forward-backward asymmetry
became insensitive to these uncertainities and therefore we do not include
them while analysing the above decay in UED model.

To see the new physics effects due to extra dimension, the differential
forward-backward asymmetry with $\hat{s}$ is plotted in Fig. 2. It can be
seen that the zero position of forward-backward asymmetry $A_{FB}$ shifts
towards the left in ACD model with single universal extra dimension and this
shifting is more clear for $1/R=200$ GeV. In future, when we have some data
on these decays, this sensitivity of the zero position to the
compactification parameter, will be used to constrain $1/R.$

The method of calculation of the form factors for $B\rightarrow K_{1}\ell
^{+}\ell ^{-}$ decay descried in \cite{24} can be straighforwardly used to
calculate the form factors for $B\rightarrow K^{\ast }\ell ^{+}\ell ^{-}$.
Now after calculating these form factors for $B\rightarrow K^{\ast }\ell
^{+}\ell ^{-}$ we have plotted the forward-backward asymmetry with $\hat{s}$
in Fig. 3. We believe that it provides a useful comparsion, if one compares
the effect of our form factors to the zero of $A_{FB}$ with the others like 
\cite{10} and references therein. Again the zero of $A_{FB}$ is shifted
towards the left in ACD model with single universal extra dimension and this
shifting is more clear for $1/R=200$ GeV.

\textbf{Conclusion}

This paper deals with the study of semileptonic decay $B\rightarrow
K_{1}\ell ^{+}\ell ^{-}$ in ACD model with single universal extra dimension
which is strong contender to study physics beyond SM and has received a lot
of interest in the literature. We studied the dependence of the physical
observables like decay rate and zero position of forward-backward asymmetry
on the inverse of compactification radious $1/R$. The value of the branching
ratio is found larger then the corresponding SM value. The zero postion of
the FB asymmetry is very sensitive to $1/R$ and it is seen that it shifts
significantly to the left. The shifting is large at $1/R=200$ GeV and
approaches to the SM value if we increase the value of $1/R$. The future
experiments, where mora data is expected, will put stringent constraints on
the compactification radius and also give us some deep understanding of $B$%
-physics.

\textbf{Figure Captions}

1): The differential branching ratio as a function of $\hat{s}$ is plotted
using the form factors defined in Eq. (\ref{form-factors}). The solid line
denotes the SM result, dashed-dotted line is for $1/R=200$ GeV and dashed
line is for $1/R=500$ GeV. All the input parameters are taken at their
central values.

2): The differential forward-backward (FB) asymmetry as a function of $\hat{s%
}$ is plotted using the form factors defined in Eq. (\ref{form-factors}).
The solid line denotes the SM result, dashed-dotted line is for $1/R=200$
GeV and dashed line is for $1/R=500$ GeV. All the input parameters are taken
at their central values.

3): The differential forward-backward (FB) asymmetry for $B\rightarrow
K^{*}\ell ^{+}\ell ^{-}$ as a function of $\hat{s}$ is plotted using the
form factors defined in Eq. (\ref{form-factors}) with obvious replacements
for $K^{*}$. The solid line denotes the SM result, dashed line is for $%
1/R=200$ GeV and long-dashed line is for $1/R=500$ GeV. All the input
parameters are taken at their central values.

4): Comparision of the differential forward-backward (FB) asymmetry for $%
B\rightarrow K^{*}l^{+}l^{-}$ in Standard Model (SM) as a function of $\hat{s%
}$ is plotted using the form factors defined in Eq.(\ref{form-factors}) vs
form factors given in Colangelo $etal$ \cite{16}. The Solid line denotes the
Colangelo result and dashed line denotes our result.

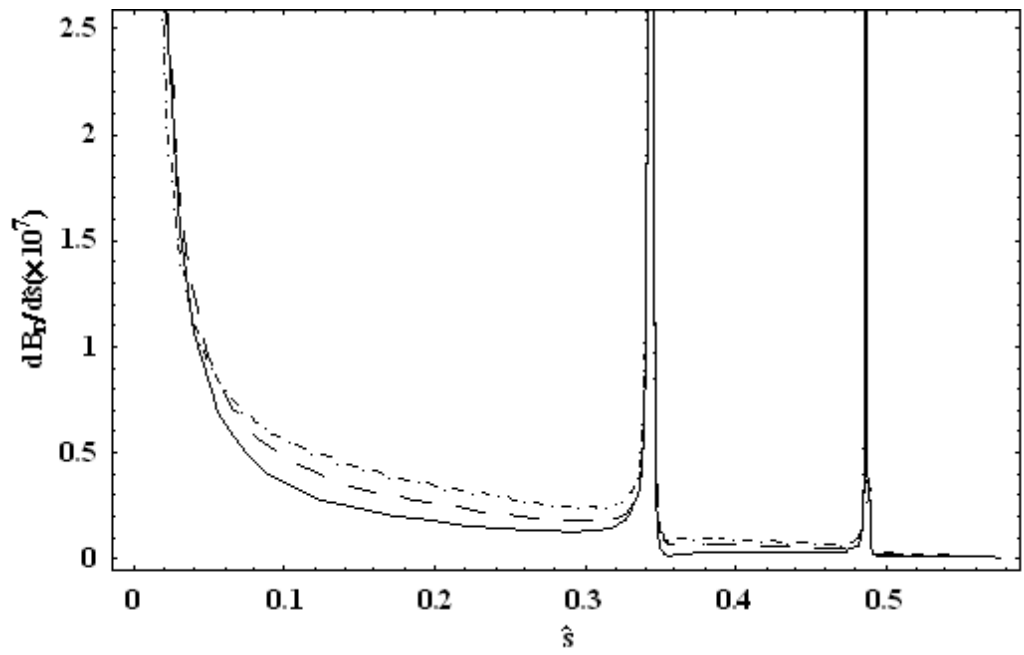
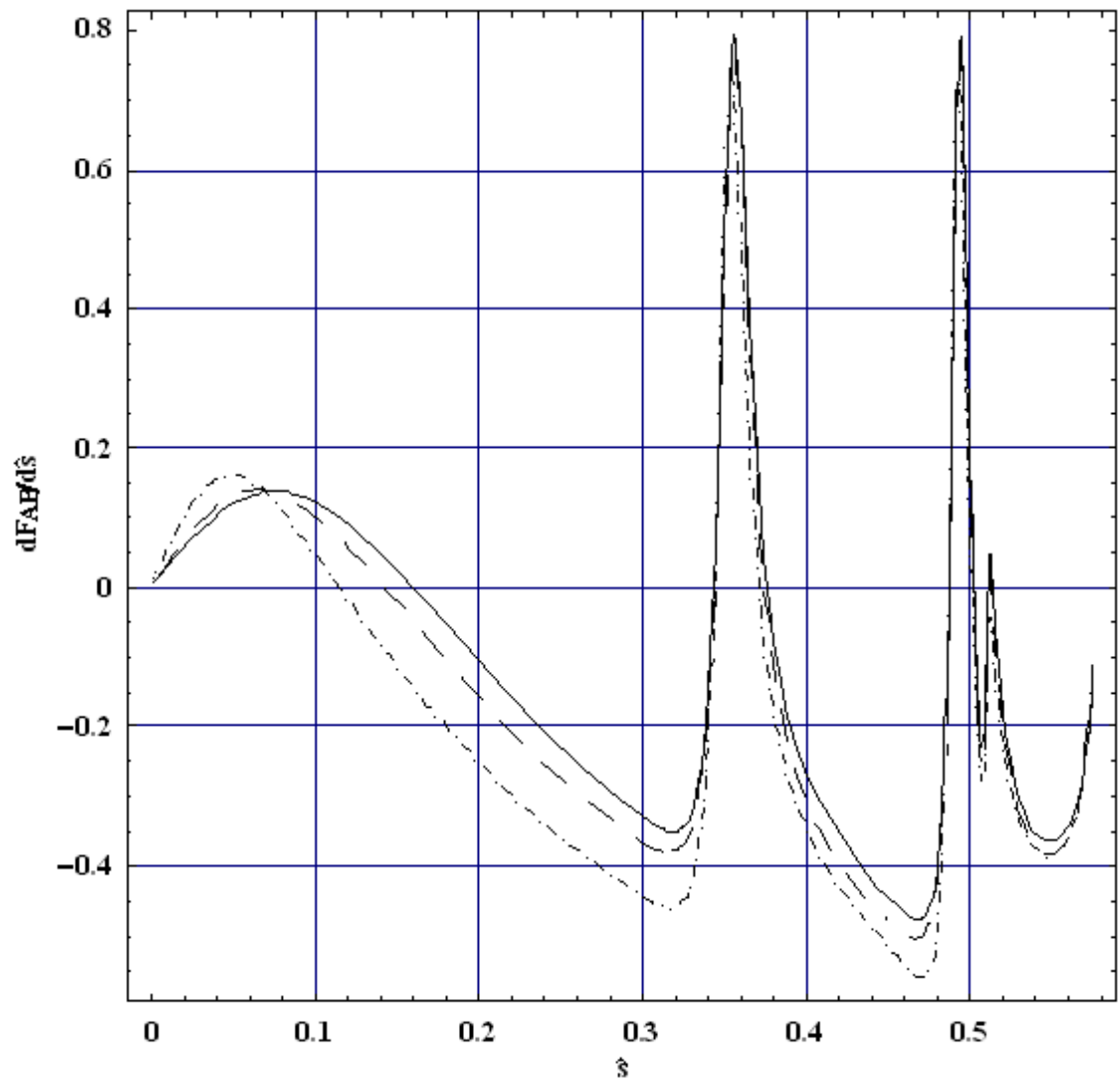
\end{document}